\documentclass[10pt]{iopart}
\usepackage[dvips]{color}
\usepackage{Graphen}
\usepackage{graphicx}
\newcommand{\arcsinh}{\rm arcsinh}

\begin{document}
\title[Hybridisation in two-band Hubbard models with different bandwidths]
{Hybridisation in Hubbard models with different bandwidths}

\author{J. B\"unemann${}^1$, D. Rasch${}^2$, and F.~Gebhard${}^1$ }

\address{${}^1$Fachbereich Physik, Philipps-Universit\"at Marburg,
D-35032 Marburg, Germany\\
${}^2$Institut f\"ur Theoretische Physik, Universit\"at zu K\"oln, 
D-50937 K\"oln, Germany }

\begin{abstract}
We investigate the orbital selective Mott transition in  
two-band Hubbard  models by means of the Gutzwiller variational theory. 
In particular, we study the influence of a finite local hybridisation 
between electrons in different orbitals on the metal-insulator
 transition.   
\end{abstract}




\section{Introduction}
\label{intro}
Metal-insulator transitions in Hubbard models with different densities of 
states have attracted a lot of interest in recent years [1-10]. 
 A dispute arose over the question whether or not
 the  transition occurs at different interaction strengths 
for the wide and the narrow band. 
A transition with different critical interaction parameters is 
usually denoted as an `orbital selective Mott transition' (OSMT).  
Apparently, a consensus has been reached that such an OSMT can occur in 
Hubbard models with different bandwidths, 
subject to the bandwidth ratio $\alpha$ of the narrow 
and the wide band and the  value of the local exchange interaction $J$.

In most of the calculations in [1-10] the dynamical mean-field theory 
has been employed. We will use multiband Gutzwiller
wave functions in order to study the OSMT. Such wave functions were 
originally introduced by Gutzwiller~\cite{Gutzwiller} 
in order to study ferromagnetism in the one-band Hubbard model. The 
evaluation of expectation values for the Gutzwiller wave function  
poses a difficult many-particle problem. Therefore, Gutzwiller, in his original work, 
used an approximation based on  quasi-classical counting arguments 
\cite{RMPVollhardt,JBcounting}. 
This `Gutzwiller approximation' later turned out to be equivalent to 
an exact evaluation of expectation values in the limit of infinite 
spatial dimension or infinite coordination number. Generalised 
Gutzwiller wave function for multi-band Hubbard models have first been 
introduced and evaluated in the limit of infinite spacial 
dimensions in reference \cite{PRB}. The formalism was further generalised, 
e.g., for superconducting  systems, in references \cite{springer,supra}. 

The OSMT in a two-band Hubbard model has first been investigated by means 
of the Gutzwiller theory in reference \cite{Fabrizio}. In that work the 
authors found an OSMT both for vanishing ($J=0$) as well as for finite 
 ($J\ne 0$) local exchange interaction. For $J=0$ the critical band width 
ratio was found to be $\alpha_c=0.2$. The Gutzwiller results in \cite{Fabrizio}
 were in good agreement with data from DMFT and a slave-spin approach 
proposed in reference \cite{Biermann}. 

In this work we will analyse the OSMT in a two-band model 
 in more detail. In particular, we permit 
 a finite expectation value  
 $\Delta_0=\langle \hat{c}_{i,1}^{\dagger} \hat{c}_{i,2} \rangle$ for the local hybridisation which can 
 change the nature of the OSMT. Such a  hybridisation could be finite 
 spontaneously, solely due to the Coulomb interaction, or due to 
 a finite hybridisation term in the Hamiltonian. 
 We will investigate both possibilities. 

Our paper is organised as follows: The two-band Hubbard models
 are introduced in section~\ref{ms}. 
In section~\ref{gwf} we define generalised Gutzwiller wave functions
 and give the results for the variational 
ground-state energy for these wave functions in the limit of infinite 
spatial dimensions. The orbital selective Mott transition in a 
 two-band model without a finite local hybridisation is discussed 
numerically, and as far as possible analytically, in section~\ref{tosmtiathm}.
 In section~\ref{tbm} we investigate analytically the spontaneous 
hybridisation in a spinless two-band model. Finally, 
the hybridisation effects in the full two-band are studied in section 
 \ref{htbm}, and a summary closes our presentation in section~\ref{sum}.

\section{Model systems}\label{ms}
\label{sec:Hamilt}
In this work we investigate the two-band Hubbard model 
\begin{equation}\label{ms1}
\hat{H}=\sum_{i,j;b;\sigma}t^{b}_{i,j}\hat{c}^{\dagger}_{i,b,\sigma}
\hat{c}^{}_{j,b,\sigma}
+\sum_{i}\hat{H}_{i;{\rm at}}=\hat{H}_0+\hat{H}_{\rm loc}\;.
\end{equation}
Here, the one particle Hamiltonian $\hat{H}_0$ describes the hopping 
of electrons with spin $\sigma$ on a lattice with  $L$ sites. 
The index $b=1,2$ labels the two degenerate orbitals at each lattice site.
We assume that the hopping amplitudes
 \begin{equation}
t^{b}_{i,j}=\alpha_{b}t_{i,j} 
\end{equation}   
 depend on the orbital index $b$ only via overall bandwidth factors 
$\alpha_{b}$. This leads to an orbital-dependent  renormalisation 
\begin{equation}\label{ms2}
  D_{b}(\varepsilon)=\frac{1}{\alpha_b}D_0 
\left(\frac{\varepsilon}{\alpha_b} \right)      
\end{equation}   
of the bare density of states
\begin{equation}\label{ms3}
D_{0}(\varepsilon)=\frac{1}{L}\sum_{k}\delta(\varepsilon-\varepsilon_{k})\;,
\end{equation}  
where $\varepsilon_{k}$ is the Fourier-transform of $t_{i,j}$. Throughout 
this work, only symmetric densities of states will be considered 
$D_{0}(-\varepsilon)=D_{0}(\varepsilon)$. 

We will study the two-band model (\ref{ms1}) with and without 
spin-degrees of freedom. For the full two-band model we assume 
that the orbitals have an $e_{\rm g}$-symmetry. The atomic Hamiltonian then 
reads
\begin{eqnarray}\label{ms5}
\hat{H}^{(2)}_{\rm{at}}&=&U\sum_{b}\hat{n}_{b,\uparrow}\hat{n}_{b,\downarrow}
+U'\sum_{\sigma,\sigma'}\hat{n}_{1,\sigma}\hat{n}_{2,\sigma'}
-J\sum_{\sigma}\hat{n}_{1,\sigma}\hat{n}_{2,\sigma}\\ \nonumber
&&-J\sum_{\sigma}\hat{c}^{\dagger}_{1,\sigma}\hat{c}_{2,-\sigma}^{\phantom{+}}\hat{c}^{\dagger}_{1,-\sigma}\hat{c}_{2,\sigma}^{\phantom{+}}
-J_C(\hat{c}^{\dagger}_{1,\uparrow}\hat{c}^{\dagger}_{1,\downarrow}\hat{c}_{2,\downarrow}^{\phantom{+}}\hat{c}_{2,\uparrow} ^{\phantom{+}}+{\rm h.c.}) 
\end{eqnarray}
where in cubic symmetry the two parameters $U'$ and $J_C$ are determined 
by $U'=U-2J$ and $J_{C}=J$.
Without spin, the atomic Hamiltonian 
$\hat{H}_{i;{\rm at}}$ simply reads
\begin{equation}\label{ms4}
\hat{H}^{(1)}_{\rm at}=U \hat{n}_{1}\hat{n}_{2}\;,
\end{equation}
where the effective Hubbard interaction in this model can 
be derived from the interorbital Coulomb ($U'$) and exchange ($J$) 
interaction through  $U=U'-J$.
Apparently, the spinless two-band model is mathematically equivalent 
to a one-band model with a spin-dependent density of states.
In the limit $\alpha_2 \rightarrow 0$ it becomes a 
Falicov-Kimball model.

Both atomic Hamiltonians (\ref{ms3}) and (\ref{ms4}) can be readily 
diagonalised
\begin{equation}\label{ms6}
\hat{H}^{(1),(2)}_{\rm at}=\sum_{\Gamma}E_{\Gamma}|\Gamma \rangle  \langle \Gamma |
\;.
\end{equation}
The eigenstates $|\Gamma \rangle$ of  $\hat{H}_{\rm at}^{(1)}$ are 
 the empty state $| \emptyset \rangle$, the two singly occupied states 
$| b \rangle$ and the doubly occupied state $| d \rangle$. 
The diagonalisation of $\hat{H}_{\rm at}^{(2)}$ leads to similar Slater-determinants 
for all particle numbers $n_{\rm at}\neq 2$. In the two-particle sector, 
$n_{\rm at}= 2$, 
one finds the triplet ground-state with energy 
$E_{\Gamma}=U-3J$, in agreement with Hund's first rule, and three 
singlet states with energies $E_{\Gamma}=U-J$ (doubly degenerate) 
and $E_{\Gamma}=U+J$; for more details, see reference \cite{PRB}. 
   
\section{Gutzwiller wave functions} \label{gwf}
\subsection{Definition}
In order to study the two-band Hubbard models introduced in 
section \ref{ms}, we use Gutzwiller 
variational wave functions \cite{Gutzwiller}  which are defined as 
\begin{equation}\label{gwf1}
| \Psi_{\rm G}\rangle \equiv \prod_{i} 
 \hat{P}_{i} | \Psi_0\rangle \; .
\end{equation}
Here, $| \Psi_0\rangle$ is a normalised one-particle wave function and the 
local correlation operator $\hat{P}_{i}$ has the form
\begin{equation}
\label{gwf2}
\hat{P}=\sum_{\Gamma,\Gamma'}\lambda_{\Gamma,\Gamma'}\hat{m}_{\Gamma,\Gamma'}\;, 
\end{equation}
for each lattice site $i$, and
\begin{equation}
\hat{m}_{\Gamma,\Gamma'}=|\Gamma \rangle  \langle \Gamma'|\;.
\end{equation}
The real coefficients $\lambda_{\Gamma,\Gamma'}$ and the one-particle 
wave function $| \Psi_0\rangle$ are variational parameters. For systems 
without superconductivity it is safe to assume that the parameters 
$\lambda_{\Gamma,\Gamma'}$ are finite only for atomic states 
$| \Gamma \rangle$, $| \Gamma' \rangle$ with the same particle number.
For ground states without spin order one can further 
assume that only states with the same $\hat{S}_{z}$ quantum number 
lead to finite non-diagonal variational parameters. Due to these symmetries 
the correlation operator (\ref{gwf2}) contains up to 5 variational parameters 
for $\hat{H}_{\rm at}^{(1)}$ and up to 26 for $\hat{H}_{\rm at}^{(2)}$. 

Throughout this work we will investigate the half-filled case of our model 
systems and allow for a finite local hybridisation
 \begin{equation}\label{gwf3}
\Delta_0=\langle \hat{c}^{\dagger}_{i,1,\sigma}
\hat{c}^{}_{i,2,\sigma} \rangle_{\Psi_0}.
 \end{equation}
With respect to the operators $\hat{c}^{\dagger}$ and $\hat{c}^{}$,
the local density matrix is therefore non-diagonal. For analytical
 and numerical calculations, it is more convenient 
to work with creation and annihilation operators
\begin{eqnarray}\label{gwf4}
\hat{h}^{(\dagger)}_{i,1,\sigma}&=&\frac{1}{\sqrt{2}}
\left(\hat{c}^{(\dagger)}_{i,1,\sigma}+\hat{c}^{(\dagger)}_{i,2,\sigma}\right)\;,\\
\hat{h}^{(\dagger)}_{i,2,\sigma}&=&\frac{1}{\sqrt{2}}
\left(\hat{c}^{(\dagger)}_{i,1,\sigma}-\hat{c}^{(\dagger)}_{i,2,\sigma}\right)
\end{eqnarray}
 which have a diagonal local density matrix,
\begin{equation}
n^{(h)}_{b}=\langle \hat{h}^{\dagger}_{i,b,\sigma}
\hat{h}^{}_{i,b',\sigma} \rangle_{\Psi_0}=\delta_{b,b'}
\left (\frac{1}{2}\pm \Delta_0 \right).
\end{equation}
With these operators the one-particle Hamiltonian $\hat{H}_0$
 reads
\begin{equation}
\hat{H}_0=\sum_{i,j;b,b';\sigma}\tilde{t}_{i,j}^{b,b'}
\hat{h}^{\dagger}_{i,b,\sigma}
\hat{h}^{}_{j,b',\sigma}
 \end{equation}
where 
\begin{equation}
\tilde{t}^{b,b'}_{i,j}=\frac{t_{i,j}}{2}(\delta_{b,b'}
+\Delta \alpha(1-\delta_{b,b'}))\;.
\end{equation}
Both atomic Hamiltonians (\ref{ms3}) and (\ref{ms4}) keep their 
form under a transformation from $\hat{c}^{}$  to $\hat{h}^{}$.
By building a basis of Slater determinants $| H\rangle$ with the 
operators $\hat{h}_{i,b,\sigma}^{\dagger}$ the eigenstates of the 
atomic Hamiltonian can be written as
\begin{equation}
| \Gamma \rangle =\sum_{H}T_{\Gamma,H} |H\rangle.
\end{equation}

 \subsection{Evaluation in infinite spatial dimensions}\label{eisd}
The evaluation of expectation values for Gutzwiller wave functions poses
a difficult many-particle problem. In this work we employ an evaluation 
scheme that becomes exact in the limit of infinite spatial dimensions. 
Within this approach the expectation value of the local Hamiltonian reads
\begin{equation}\label{eisd1}
\langle \hat{H}_{\rm at} \rangle_{\Psi_{\rm G}}
=\sum_{\Gamma,\Gamma_1,\Gamma_2} E_{\Gamma}
 \lambda_{\Gamma_1,\Gamma} \lambda_{\Gamma,\Gamma_2} 
\langle \hat{m}_{\Gamma_1,\Gamma_2}   \rangle_{\Psi_{0}} \;.
\end{equation}
Here, the expectation value 
$\langle \hat{m}_{\Gamma_1,\Gamma_2}   \rangle_{\Psi_{0}}$ is given as
\begin{equation}\label{eisd2}
\langle \hat{m}_{\Gamma_1,\Gamma_2}   \rangle_{\Psi_{0}}
=\sum_{H}T_{\Gamma_1,H}T_{\Gamma_2,H}m^{0}_{H}
\end{equation}
where
\begin{equation}\label{eisd3}
m^{0}_{H}=\prod_{b({\rm occ.})}n^{(h)}_{b}\prod_{b({\rm unocc.})}(1-n^{(h)}_{b})\,.
\end{equation}
For the expectation value of a hopping term in the one-particle 
Hamiltonian one finds
\begin{equation}\label{eisd4}
\langle \hat{h}^{\dagger}_{i,b,\sigma}
\hat{h}^{}_{j,b',\sigma} \rangle_{\Psi_{\rm G}}=
\sum_{\tilde{b},\tilde{b'}}\tilde{q}_{b \tilde{b}}\tilde{q}_{b' \tilde{b}'}
\langle \hat{h}^{\dagger}_{i,b,\sigma}
\hat{h}^{}_{j,b',\sigma} \rangle_{\Psi_{0}}\;,
\end{equation}
where the elements of the renormalisation matrix $\tilde{q}$ are given as
\begin{equation}\label{eisd5}
q_{b \tilde{b}}=\sum_{\Gamma_1,\Gamma_2,\Gamma_3,\Gamma_4}
\lambda_{\Gamma_1,\Gamma_2}\lambda_{\Gamma_3,\Gamma_4}
\langle  \Gamma_2 |\hat{h}^{\dagger}_{i,b,\sigma} | \Gamma_3 \rangle
\frac{
\left
\langle 
\left(
\hat{h}^{\dagger}_{i,\tilde{b},\sigma} |  \Gamma_4 \rangle  \langle 
\Gamma_1|
\right)
\right
\rangle_{\Psi_0}}
{1-n^{(h)}_{\tilde{b}}}.
\end{equation}
The remaining expectation value in (\ref{eisd5}) can be calculated 
in the same way as (\ref{eisd2}). Note the symmetries 
$\tilde{q}_{1,1}=\tilde{q}_{2,2}$ and $\tilde{q}_{1,2}=\tilde{q}_{2,1}$. 
The renormalisation factors for the 
$\hat{c}$-operators are diagonal,
\begin{equation}\label{eisd4b}
\langle \hat{c}^{\dagger}_{i,b,\sigma}
\hat{c}^{}_{j,b,\sigma} \rangle_{\Psi_{\rm G}}=
q_{b}^2
\langle \hat{c}^{\dagger}_{i,b,\sigma}
\hat{c}^{}_{j,b,\sigma} \rangle_{\Psi_{0}}
\end{equation}
and given by 
\begin{equation}
q_{{}_{\left(1\atop{2}\right)}}=\tilde{q}_{1,1}\pm\tilde{q}_{1,2}\,.
\end{equation}

Furthermore, the evaluation in infinite dimensions shows that the variational 
parameters $\lambda_{\Gamma,\Gamma'}$ and the one-particle wave function 
 $|\psi_0 \rangle$ have to obey the constraints
\begin{equation}\label{eisd6}
1=\langle \hat{P}^{2}  \rangle_{\psi_0}=
\sum_{\Gamma,\Gamma_1,\Gamma_2}
 \lambda_{\Gamma_1,\Gamma} \lambda_{\Gamma,\Gamma_2} 
\langle \hat{m}_{\Gamma_1,\Gamma_2}   \rangle_{\Psi_{0}} 
\end{equation}
and
\begin{eqnarray}\label{eisd7}
n^{(h)}_{b}\delta_{b,b'}&=& \langle \hat{P}^{2} 
\hat{h}^{\dagger}_{b,\sigma}  \hat{h}^{}_{b',\sigma}  
\rangle_{\psi_0} \\ \nonumber
&=&\sum_{\Gamma,\Gamma_1,\Gamma_2}
 \lambda_{\Gamma_1,\Gamma} \lambda_{\Gamma,\Gamma_2} 
\langle \hat{m}_{\Gamma_1,\Gamma_2}  
\hat{h}^{\dagger}_{b,\sigma}  \hat{h}^{}_{b',\sigma}  
 \rangle_{\Psi_{0}} \;.
\end{eqnarray}

\section{The orbital selective Mott transition in a two-band Hubbard model}\label{tosmtiathm}
\begin{figure}[ht]
\includegraphics[clip,height=4.5cm]{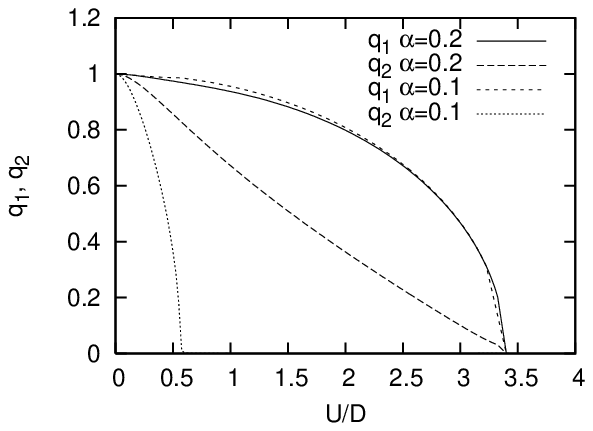}
\includegraphics[clip,height=4.5cm]{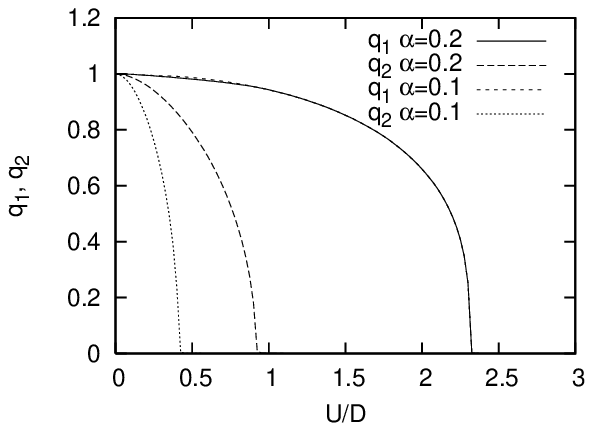}
\caption{Renormlisation factors $q_{b}$ for $\Delta=0$,
and bandwidth ratios $\alpha=0.2$, $\alpha=0.1$; \textit{left}: $J=0$
; \textit{right}: $J=0.1$}.\label{fig4a}
\end{figure}

In this section we investigate the metal-insulator transition 
 in the two-band Hubbard model without local hybridisation. 
 We use a semi-elliptic density of states 
\begin{equation}
D_{0}(\varepsilon)=\frac{2}{\pi}\sqrt{1-\varepsilon^2}
\end{equation}  
which leads to the bare one-particle energy
\begin{equation}\label{tbm2}
\varepsilon_0=\int_{-\infty}^{0}d\varepsilon  D_0(\varepsilon ) \varepsilon
=-\frac{2}{3\pi}\;.
\end{equation}
Our energy unit is given by $D=1$, half of the bare bandwidth.  
When we set $\alpha_1=1$ and introduce the bandwidth ratio 
$\alpha\equiv \alpha_1/\alpha_2\leq 1$,
the expectation value for the one-particle Hamiltonian in (\ref{ms1}) 
is given as
\begin{equation}
\langle\hat{H}_0\rangle_{\Psi_{\rm G}}=(q_1^2+q_2^2 \alpha)\varepsilon_0\,.
\end{equation}
Without hybridisation, the variational ground-state energy has to be 
minimised only with respect to the variational parameters 
$\lambda_{\Gamma,\Gamma'}$.
In figure \ref{fig4a} (left) we show the resulting renormalisation factors
 $q_b$ as a function of $U$ for $J=0$ and two different bandwidth ratios 
$\alpha$. As already observed in reference \cite{Fabrizio}, 
it depends on the value of $\alpha$ 
whether or not there is an orbital selective Mott transition. For $J=0$, the 
critical ratio is $\alpha_{\rm c}=0.2$, i.e., the renormalisation factors 
$q_{1}$, $q_{2}$  vanish at two different critical values 
$U_{\rm c2}<U_{\rm c1}$ if $\alpha<\alpha_{\rm c}$.
By switching on $J$, the critical ratio $\alpha_{\rm c}$ becomes larger and 
the Mott transitions take place at smaller values of $U$; see figure \ref{fig4a}(right). 

\begin{figure}[ht]
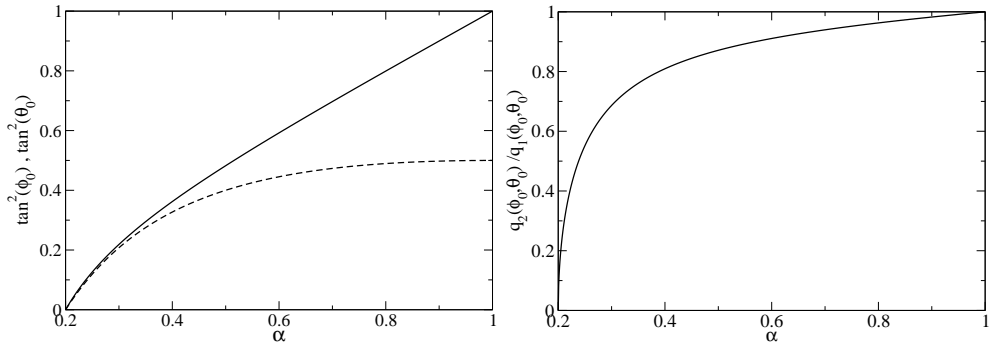

\includegraphics[clip,height=4.5cm]{phi_theta.eps}
\includegraphics[clip,height=4.5cm]{qrat.eps}
\caption{\textit{left}: $\tan{(\phi_0)}^2$ (straight) and 
$\tan{(\theta_0)}^2$ (dashed) 
at the Mott transition as a function of bandwidth ratio $\alpha$;
\textit{right}: Ratio of renormalisation factors $q_2/q_1$ at the Mott 
transition  as a function of $\alpha$.}\label{fig4b}
\end{figure}

For $J=0$, we can gain more insight into the nature of the 
different Mott-transitions in our model by some analytical calculations.
First, we consider the case $\alpha>\alpha_{\rm c}$. If we approach the 
Mott transition from below, we can neglect the variational parameters 
$m_{\emptyset}=m_{4}$ for empty and fourfold occupied sites. Due to the 
 high symmetry of the model for $J=0$ the ground-state energy is then 
a function of only three variational parameters, $d$, $\phi$, and, $\theta$,
\begin{equation}
E=2\varepsilon_0 d\left(1-2d\right)f\left(\phi,\theta\right)+\left(1+d \right)U
\end{equation}
where
\begin{eqnarray}
f\left(\phi,\theta\right)&=&4\alpha_1\left(\sin\left(\phi\right)\sin\left(\theta\right)+
\sqrt{2}\cos\left(\phi\right)\cos\left(\theta\right)\right)^2\\
&&+4\alpha_2\left(\cos\left(\phi\right)\sin\left(\theta\right)+\sqrt{2}\sin\left(\phi\right)\cos\left(\theta\right)\right)^2.
\end{eqnarray}
 Here, $\tan{(\phi)}^2$ gives the ratio of the probabilities  to
 find a singly occupied site with an electron in the wide and in the narrow 
orbital. The ratio of the  probabilities for doubly occupied sites with two 
electrons in the same and in different orbitals is parametrized by 
$\tan{(\theta)}^2$. The variational parameter $d$ gives the total probability 
for single occupation. At the Mott transition, where $d\rightarrow 0$, the two 
angles $\phi$, $\theta$ can be calculated analytically 
\begin{eqnarray}
\theta_0&\equiv&\theta(d\rightarrow 0)=\frac{1}{2}\arccos\left(\frac{-17+2\alpha-17 \alpha^2}{3\left(1-34\alpha+\alpha^2\right)}\right) \,,\\
\phi_0&\equiv&\phi(d\rightarrow 0)=\frac{1}{2}\arctan{(\frac{(1+\alpha)2 \sqrt{2}\sin{2\theta_0}}
{(1-\alpha)(1+\cos{2\theta_0})})}\,.
\end{eqnarray}     
 Both values, $\tan{(\phi_0)}^2$, and  $\tan{(\theta_0)}^2$  are shown as a
 function of $\alpha$ in figure \ref{fig4b}(left).

 As expected, the weight of local states with no electron in the 
 narrow band vanishes for $\alpha \rightarrow \alpha_{\rm c}$. The 
renormalisation factors $q_b$ both vanish proportional to a square-root,  
$q_b\sim\sqrt{U_{c}-U}$, when $U$ approaches $U_{c}$ from below.
The ratio $q_2/q_1$ is finite for $U\rightarrow U_{c}$ and goes to zero 
proportional to $\sqrt{\alpha-\alpha_{c}}$, see figure \ref{fig4b} (right). 
Finally, the critical interaction strength $U_{{\rm c}2}=U_{{\rm c}1}$ is given 
as
\begin{equation} \label{uc1a}
U_{{\rm c} 1}=2|\varepsilon_0|f(\phi_0,\theta_0) \,\,\,
\,\,\,\,\,\,(\alpha>\alpha_{\rm c})\;\;.
\end{equation}

\begin{figure}[ht]
\centerline{
\includegraphics[clip,height=5.5cm]{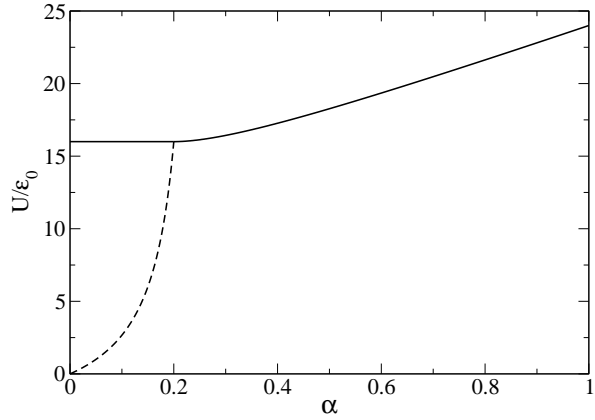}}
\caption{Critical interaction parameters $U_{{\rm c}1}$ (straight) 
and $U_{{\rm c}2}$ (dashed) as a function of $\alpha$ (see eqs. 
(\ref{uc1a}), (\ref{uc1b}), (\ref{uc2a})).}\label{fig4c}
\end{figure}

Next, we consider the case $\alpha<\alpha_{\rm c}$. For interaction parameters
$U_{{\rm c} 2}<U<U_{{\rm c}1}$, the electrons in the narrow band are localised 
and  the wide band can be treated as an effective one-band model. 
This leads us to the critical interaction parameter 
\begin{equation}  \label{uc1b}
U_{{\rm c}1}=2|\varepsilon_0|f(0,0)=16 |\varepsilon_0| \,\,\,
\,\,\,\,\,\,(\alpha<\alpha_{\rm c})  
\end{equation} 
for the Brinkmann-Rice transition of the wide band. Starting from the 
Brinkmann-Rice solution for $U<U_{{\rm c}1}$, we can expand the 
variational energy to leading (i.e. second) order with respect to the 
three parameters $\{v_i\}=\{\phi,\theta,m_{\emptyset}\}$,
\begin{equation} 
E=E_0+\sum_{i,j=1}^{3}v_i\tilde{E}_{i,j}v_j\,.
\end{equation} 
 The localisation of the narrow band becomes unstable when the matrix 
$\tilde{E}$ has negative eigenvalues for physical parameters $v_i>0$. 
 This evaluation yields the following expression for the narrow-band 
critical interaction strength
\begin{equation}  \label{uc2a}
U_{{\rm c}2}=16 |\varepsilon| \frac{\alpha}{1-4\alpha}\,\,\,\,
\,\,\,\,\,\,(\alpha<\alpha_{\rm c}) .
\end{equation} 
The resulting phase diagram for all $0\leq \alpha\leq1 $  is shown in
 figure \ref{fig4c}. 

\section{The spinless two-band model}
\label{tbm}

As the simplest example for a model with different densities of states 
 we investigate the spinless two-band model.
 In the half filled case and without spontaneous hybridisation ($\Delta_0=0$)
the constraints  (\ref{eisd6}) and (\ref{eisd7}) can be solved analytically 
for this model. The variational energy is then solely a 
function of $\lambda_d$,
\begin{equation}\label{tbm1}
E_{\rm var}=4 \lambda_d^2 \left(1-\frac{\lambda_d^2}{2}\right)
 \varepsilon_0+\frac{U}{4} \lambda_d^2 \;.
\end{equation}
The energy (\ref{tbm1}) can be minimised analytically. As a result
 one finds the well known Brinkmann-Rice solution 
\begin{eqnarray}\label{tbm3}
q_{\rm BR}&=&1-\left(\frac{U}{U_{\rm c}} \right)^2 \,,\\
d_{\rm BR}&=&\frac{1}{4}\left(1-\frac{U}{U_{\rm c}} \right)
\end{eqnarray}
for the renormalisation factor $q=\delta_{b,b'}q_{b,b'}$ and the expectation
value of the double occupancy $d=\lambda_d^2/4$. The Brinkmann-Rice 
metal insulator transition occurs at the critical value 
$U=U_{\rm c}\equiv 16|\varepsilon_0|$. 

For the renormalisation factors
$\alpha_{b}$ we set $\alpha_1+\alpha_2=2$, i.e. the difference 
of the bandwidths is parametrized by  
$\Delta \alpha\equiv \alpha_1-\alpha_2$.
Starting from the analytic solution for vanishing hybridisation
we can calculate the variational ground state energy to leading order 
in $\Delta_0$, 
\begin{equation}\label{tbm4}
E_{\Delta_0}=E_{\rm BR}+C(U,\Delta \alpha) \Delta_0^2 \;.
\end{equation}
 A spontaneous hybridisation  will appear if the 
coefficient $C$ in (\ref{tbm4}) is negative. The analytical evaluation 
 leads to the Stoner-type instability criterion 
 \begin{equation}\label{tbm5}
\frac{f(\Delta \alpha)}
{U_{\rm c}D_0(0)}<
\frac{
U/U_{\rm c}
\left( 
2+U/U_{\rm c} 
\right)}
{2
\left(
1+U/U_{\rm c}
\right)^2}\equiv g\left(U/U_{\rm c} \right),
\end{equation}
where
\begin{equation}\label{tbm6}
f(\Delta \alpha)\equiv \frac{\Delta \alpha}
{2 \arcsinh\left(
\Delta \alpha/\sqrt{4-\Delta \alpha^2}
\right)}.
\end{equation}
\begin{figure}[ht]\label{fig5a}
\includegraphics[clip,height=4.5cm]{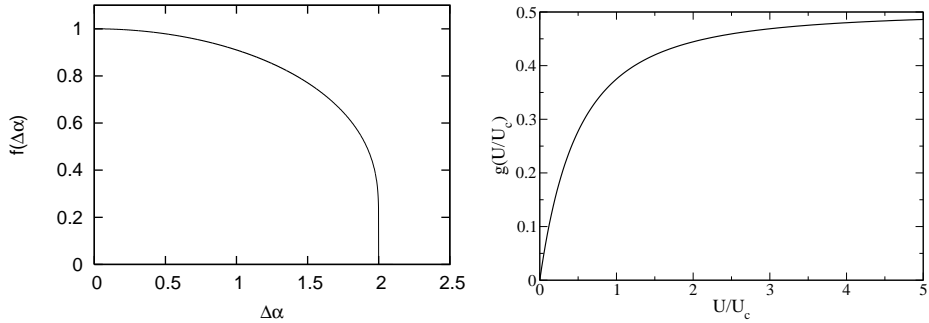}
\includegraphics[clip,height=4.2cm]{stonercrit.eps}
\caption{\textit{left}: $f\left(\Delta\alpha\right)$;
\textit{right}: $g(U/U_{\rm c})$}
\end{figure}
In figure \ref{fig5a} the function $f(\Delta \alpha)$ and the right
 hand side of equation (\ref{tbm5}) are shown as a function of $\Delta \alpha$
 and $U$, respectively. As can be seen from this figure the function 
 $f(\Delta \alpha)$ and therefore the left hand side of (\ref{tbm5})
 approach zero for $\Delta \alpha \rightarrow 2$. 
 On the other hand, the right hand side of (\ref{tbm5}) is positive for all 
 $U>0$. This means that for arbitrary 
values of $U$ there exist a critical bandwidth difference 
$\Delta \alpha_{\rm c}$ with $\Delta_0>0$ for $\alpha> \alpha_{\rm c}$.
Figure \ref{fig5c} (left) shows the phase-diagram for ground states with and 
without finite hybridisation for different values of the density of 
states $D_0(0)$ at the Fermi-level. Whether or not there is a transition 
in the large $U$ limit for all values of $\Delta \alpha$ depends 
on the value of $D_0(0)$. 
This is illustrated in figure \ref{fig5c} (right) where the critical difference 
$\Delta \alpha_{\rm c}$  for the transition is shown as a 
function of $D_0(0)$ in the limit $U\rightarrow \infty$. 
Note that a spontaneous hybridisation 
has already been observed in a Falicov-Kimball model within a mean-field 
 approximation \cite{czycholl}. This is in agreement with our results in the limit 
$\Delta \alpha \rightarrow 2$. 
\begin{figure}[ht]\label{fig5c}
\includegraphics[clip,height=4.5cm]{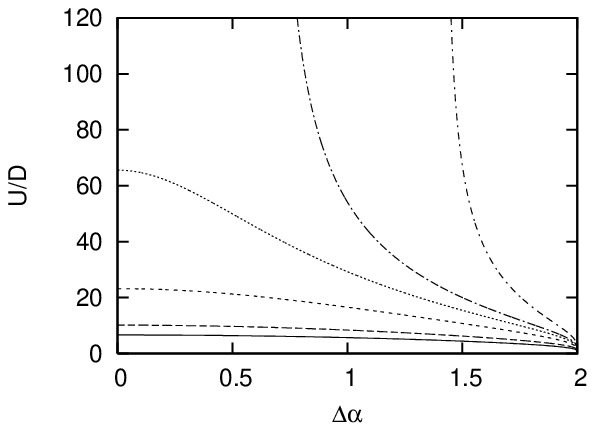}
\includegraphics[clip,height=4.5cm]{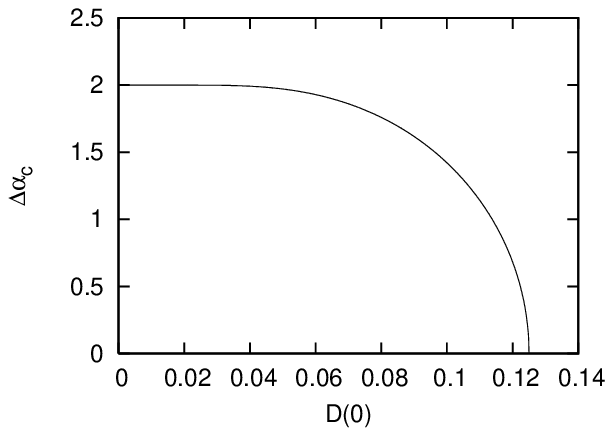}
\caption{\textit{left}:phase diagram of the spinless two-band Hubbard model
 for different densities of states at the Fermi level 
$D(0)=0.25, 0.2,0.15,0.13,0.125,0.1$ (from the bottom to the top of the figure)
\textit{right}: critical difference 
$\Delta \alpha_{\rm c}$  in the limit $U\rightarrow \infty$ as a function 
of $D_0(0)$.}
\end{figure}

In summary, our analytical results on the spinless two-band Hubbard model 
show that a difference in the bandwidth increases the tendency of the
 system to exhibit spontaneous hybridisation between the narrow band and the 
wide band. Mathematically, the reason for this is quite simple. 
Both, the expectation value of the one-particle energy 
$\hat{H}_0$ and the  Coulomb interaction $\hat{H}_{\rm loc}$  are changing 
quadratically in $\Delta_0$. However, in the limit 
$\Delta \alpha \rightarrow 0$ the energy gain from $\hat{H}_{\rm loc}$ 
always beats the rise in energy due to 
$\hat{H}_0$. At first glance, one might think that the same behaviour 
should be observed in the OSMT phase of the two-band model with the only
 difference that it is not the
bare but the effective width of the narrow band that vanishes. As we will 
discuss in the next section, however, this hypothesis turns out 
to be incorrect.

\section{Hybridisation in the two-band model}
\label{htbm}
 In this section we present numerical results for the two-band model with 
a finite local hybridisation (\ref{gwf3}). The hybridisation can develop 
 either spontaneously, like in the spinless model (section \ref{tbm}), or it 
 can be caused by a finite hybridisation term in the Hamiltonian. We will 
 discuss both effects separately.

 \subsection{Spontaneous hybridisation}
\begin{figure}[t]
\centerline{
\includegraphics[clip,height=5.5cm]{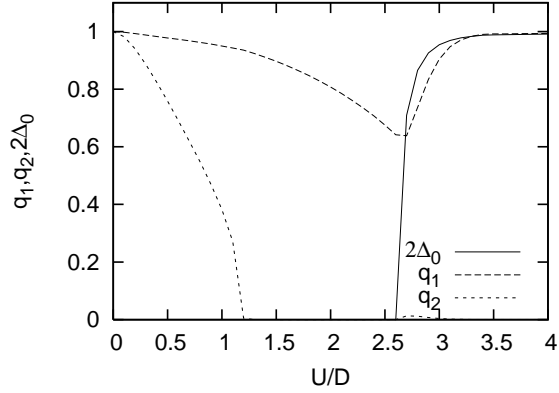}}
\caption{Renormalisation factors $q_1$, $q_2$ and hybridisation
$2\Delta_0$ for $J=0$ and $\alpha=0.15$.  }\label{fig7a}
\end{figure}
As shown in section \ref{tbm}, a vanishing width of the narrow band
 can be the driving force for a spontaneous local hybridisation of the
 wide and the narrow band. In our two-band model, however, the vanishing 
of the effective bandwidth for $q_2\rightarrow 0$  does not have the same 
effect. This can be seen in figure \ref{fig7a}, where we show the results 
for the renormalisation factors $q_1$, $q_2$ and the hybridisation 
$\Delta_0$. Unlike in the spinless model, there is 
not necessarily a finite hybridisation if the effective narrow bandwidth
goes to zero for $U\rightarrow U_{{\rm c}2}$. The reason for this 
 differing behaviour is an additional contribution to the 
 one-particle energy of the full two-band model. To leading order 
in $\Delta_0$ there is a third  term from the expansion of the 
narrow-band renormalisation factor
\begin{equation}
 q_{2}\approx  q_{2}(\Delta_0=0)+c \Delta_0^2\,.
\end{equation}
 The coefficient $c$ is negative and, multiplied with the negative 
 bare one-particle energy of the narrow band it leads to an increase of 
 the total energy. This contribution to the energy overcompensates the
 negative term from the Coulomb interaction. 

A finite  hybridisation $\Delta_0$ sets in at larger values of $U$ when the 
system is already in the OSMT phase, see figure \ref{fig7a}. 
Numerically, it seems as if $\Delta_0$ approaches its maximum 
value $\Delta_0^{\rm max}=1/2$ only in the limit $U\rightarrow \infty $. 

In all systems with finite values of $J$ that we investigated, we did not find 
a solution with spontaneous hybridisation. It is more likely, though,
 that for values of $J$ smaller than some critical parameter $J_{\rm c}$
 there is a solution with a finite hybridisation. However, it is difficult 
to determine this small parameter $J_{\rm c}$ numerically.    

 \subsection{Finite hybridisation in the Hamiltonian}   
The assumption that there is no hybridisation between the two degenerate 
bands in the Hamiltonian of our model is quite artificial. In this 
section we will therefore investigate how the OSMT is affected if we add 
a hybridisation term of the form
\begin{equation}
\hat{H}_{\rm hyb}=-\tilde{\eta}\sum_{i,\sigma}\hat{c}^{\dagger}_{i 1\sigma}\hat{c}_{i 2\sigma}+{\rm h.c.}
\end{equation}
to our Hamiltonian (\ref{ms5}). 
\begin{figure}[t]
\includegraphics[clip,height=4.7cm]{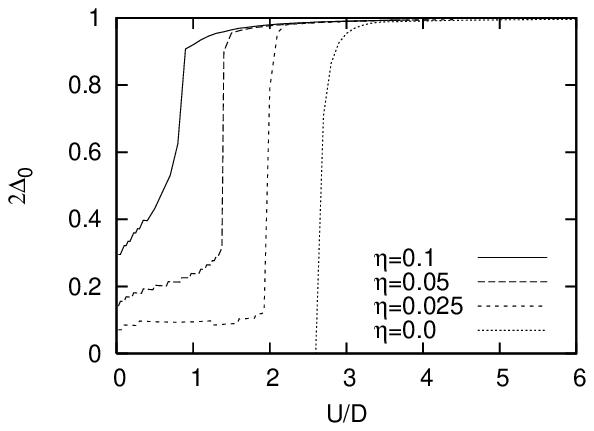}
\includegraphics[clip,height=4.3cm]{qfac.eps}
\caption{\textit{left}:expectation value $2\Delta_0$ as a function of $U$
 for several values of $\tilde{\eta}$;
\textit{right}:Renormalisation factors $q_1$, $q_2$ for 
$\alpha=0.15$ and $\tilde{\eta}=0.025 D$,  $J=0.05U$ (straight), 
 $\tilde{\eta}=0.025 D$,  $J=0.025U$ (dashed), 
$\tilde{\eta}=0.05 D$,  $J=0.025U$ (dotted). }\label{fig8}
\end{figure}
For $J=0$ we find that the OSMT phase is destroyed for any finite 
value of $\tilde{\eta}$. This is illustrated in figure \ref{fig8} (left) 
where we show the expectation value $\Delta_0$ as a function of $U$
 for several values of $\tilde{\eta}$.

For finite $J$, the behaviour of our model is more ambiguous. 
As we have seen before, a finite $J$ stabilizes the OSMT phase 
whereas a finite $\tilde{\eta}$ tends to destroy it. Therefore, it depends 
on the ratio of both quantities whether or not an OSMT is found.
Figure \ref{fig8} (right) shows the renormalisation factors $q_b$ for different 
values of $J$ and $\tilde{\eta}$. For $J=0.025U$ and 
$\tilde{\eta}=0.05 D$ the OSMT is completely suppressed. 
This is still the case for the smaller value $\tilde{\eta}=0.025 D$,
although the narrow band factor $q_2$ is already quite small in the region
of $U$ parameters where it would be zero for $\tilde{\eta}=0$. 
Finally, for larger values $J=0.05U$ an OSMT phase is restored for 
 interaction parameters $U>U_{{\rm c}2}$ where $U_{{\rm c}2}$ is larger then 
the corresponding value for $\tilde{\eta}=0$.

In summary, our numerical calculations show that appearance and 
disappearance of an OSMT results from a subtle interplay of the local 
exchange interaction $J$  and the local hybridisation $\tilde{\eta}$.  

\section{Summary}\label{sum}
In this work we have investigated the orbital selective Mott transition (OSMT)
 in two-band Hubbard models with different densities of states 
by means of the 
 Gutzwiller variational theory. We were particularly interested in 
 the question how the OSMT is modified when we allow for a finite local
 hybridisation between the wide band and the narrow band. In the two-band model 
 without spin-degrees of freedom there always is a spontaneous 
 hybridisation if the narrow bandwidth goes to zero. However, we did not find
 such a behaviour in the full two-band model. There, spontaneous hybridisation 
 was only seen for vanishing local exchange interaction, $J=0$, and for 
 Coulomb parameters $U$ larger then the critical parameter at which 
 the electrons in the narrow band localise. By adding a local  hybridisation term 
$\sim \tilde{\eta}$  to the 
Hamiltonian, the phase diagram becomes more involved. Whether or not
 an OSMT takes place depends on the relative strength of $J$ and 
$\tilde{\eta}$. The exchange interaction $J$ tends to stabilise the 
 OSMT phase, whereas the hybridisation $\tilde{\eta}$ tends to destroy it.

\ack


\end{document}